\documentclass[aps,pra,superscriptaddress,twocolumn]{revtex4-1} 

\usepackage{amsmath}
\usepackage{mathrsfs}
\usepackage{epstopdf}
\usepackage{graphicx}
\usepackage{bm,color,bbm}
\usepackage{hyperref, mathtools}
\usepackage{caption}
\usepackage{subcaption}

\newcommand{\nn}{\nonumber}

\begin{document}

\preprint{}

\title{Position-Momentum Bell-Nonlocality with Entangled Photon Pairs}


\author{James Schneeloch}
\affiliation{Department of Physics and Astronomy, University of Rochester, Rochester, NY 14627}
\affiliation{Center for Coherence and Quantum Optics, University of Rochester, Rochester, New York 14627, USA}
\affiliation{Air Force Research Laboratory, Information Directorate, Rome, NY 13441, USA}
\author{Samuel H. Knarr}
\affiliation{Department of Physics and Astronomy, University of Rochester, Rochester, NY 14627}
\affiliation{Center for Coherence and Quantum Optics, University of Rochester, Rochester, New York 14627, USA}
\author{Daniel J. Lum}
\affiliation{Department of Physics and Astronomy, University of Rochester, Rochester, NY 14627}
\affiliation{Center for Coherence and Quantum Optics, University of Rochester, Rochester, New York 14627, USA}

\author{John C. Howell}
\affiliation{Department of Physics and Astronomy, University of Rochester, Rochester, NY 14627}
\affiliation{Center for Coherence and Quantum Optics, University of Rochester, Rochester, New York 14627, USA}


\date{\today}

\begin{abstract}
Witnessing continuous-variable Bell nonlocality is a challenging endeavor, but Bell himself showed how one might demonstrate this nonlocality. Though Bell nearly showed a violation using the CHSH inequality with sign-binned position-momentum statistics of entangled pairs of particles measured at different times, his demonstration is subject to approximations not realizable in a laboratory setting. Moreover, he doesn't give a quantitative estimation of the maximum achievable violation for the wavefunction he considers. In this article, we show how his strategy can be reimagined using the transverse positions and momenta of entangled photon pairs measured at different propagation distances, and we find that the maximum achievable violation for the state he considers is actually very small relative to the upper limit of $2\sqrt{2}$.  Although Bell's wavefunction does not produce a large violation of the CHSH inequality, other states may yet do so.
\end{abstract}


\maketitle


\section{Introduction}
Though there are many predictions of quantum mechanics that disagree with classical intuition, perhaps the most striking finding is that quantum mechanics predicts the violation of Bell inequalities, ruling out the possibility that correlations between distant events can always be explained by shared information in the past. Violating Bell inequalities is more than experimental metaphysics; Bell-nonlocal entangled systems (i.e., those whose statistics cannot be described by a Local Hidden Variable (LHV) model) can be used to prove secure key rates in device-independent quantum key distribution \cite{Acin2007}. Of particular significance in quantum information protocols is the demonstration of Bell nonlocality in continuous-variable (CV) systems; their high dimensionality offers the possibility of transmitting much more information with individual particles than what their spins or polarizations can convey.

To demonstrate Bell-nonlocality \cite{BrunnerBellRMP2014} in continuous observables, one must violate a continuous-variable Bell inequality. By some standards, a fully general (i.e., contextual) continuous-variable Bell inequality does not yet exist, since we cannot deduce Bell nonlocality (yet) simply from knowing the first and higher-order moments of continuous observables for different measurement settings. However, there has been much research into demonstrating continuous-variable nonlocality by examining low-dimensional observables derived from the continuous observables of interest (examples include pseudo-spin observables \cite{ChenBellPseudoSpinPRL2002} and parity observables \cite{AbouraddyParityBellPRA2007}). In addition, there have been efforts in demonstrating continuous-variable Bell nonlocality with the statistics of discrete functions (e.g., binning functions) of continuous observables \cite{MunroBellHomodynePRA1999, WegnerBellHomodynePRA2003, NhaCVHomodyneBellPRL2004, PatronBellHomodynePRL2004, PatronHomodyneBellPRA2005, AcinBellHomoPRA2009} (in these cases, field quadratures). Indeed, since an LHV model for a continuous-variable joint probability distribution is also a model for its consequent statistics, violating the CHSH inequality with such (e.g., discrete) statistics will demonstrate the nonlocality of the underlying continuous observables.

In fact, one of the first historical attempts at showing that continuous-variable nonlocality was possible was developed by Bell himself \cite{bell1986epr, bell2004speakable}, who showed that with sign-binning, there were wavefunctions that may violate the Clauser-Horne-Shimony-Holt (CHSH) inequality \cite{CHSHbell1969}. Ironically, he also showed that the position-momentum statistics of the maximally entangled Einstein-Podolsky-Rosen (EPR) state cannot violate a Bell inequality, since they admit an explicit LHV model.

In spite of theoretical demonstrations that it is possible to demonstrate CV Bell nonlocality with sign binning, there have been no experiments to date that give a successful demonstration of Bell nonlocality for continuous observables (though there has been progress in doing so for field quadratures \cite{MunroBellHomodynePRA1999, WegnerBellHomodynePRA2003, NhaCVHomodyneBellPRL2004, PatronBellHomodynePRL2004, PatronHomodyneBellPRA2005, AcinBellHomoPRA2009}). In this article, we show how one might plausibly demonstrate CV Bell nonlocality with the transverse spatial statistics of entangled photon pairs. Futhermore, we show how the nonlocal state Bell considers actually gives a very small violation of the CHSH inequality.

To those familiar with the transverse spatial statistics of highly entangled photon pairs, it may seem impossible to demonstrate nonlocality in that degree of freedom. Indeed, it is reasonably popular to approximate the joint transverse spatial amplitude of such photon pairs as a Double-Gaussian \cite{LawEberly2004,Fedorov2009} function. Such wavefunctions, have Gaussian Wigner functions, which are non negative, and so admit an LHV model (as shown later). However, there are multiple states of entangled photon pairs (see Fig.~1) whose Wigner functions have significant regions of negative values. Though this does not, by itself guarantee nonlocality, it provides sufficient motivation to see if there are states other than the one Bell considers that might give a more substantial violation.

\subsection{Continuous-Variable CHSH Inequality with Sign Binning}
In a narrow sense, all Bell inequalities descend in one form or another from the same model of contextual \footnote{All LHV models are considered contextual in that one's measurement outcomes depend (if only trivially) on the measurement context (i.e., the known and unknown circumstances of the experimental apparatus) in addition to known and unknown factors about the system to be measured. However, there is a subset of LHV models in which the dependence of measurement outcomes on the aspects of the experimental setup is trivial (i.e. that the outcomes are independent of the measurement context). Because these \emph{non-contextual} LHV models form a subset of contextual LHV models, violating an inequality formed from non-contextual LHV models (such as those with an underlying joint probability distribution of all measurement outcomes) doesn't rule out all LHV models.} Local Hidden Variables (LHVs) that gave rise to the original CHSH inequality \cite{CHSHbell1969}. As such, many current tests of Bell nonlocality rely on one form or another of the CHSH inequality \footnote{Note: the LHV model from which the CHSH inequality is derived is also the one from which the CH76 inequality is derived.}. 

To begin, an LHV model for, say, the measurement outcomes of the positions $x_{1}$ and $x_{2}$ of a pair of particles is one where the correlations between $x_{1}$ and $x_{2}$ are explained (if not also determined) by a complete knowledge of every piece of information that could possibly be conveyed to each particle (at or below the speed of light) from points in the past. To describe these pieces of information in a general way, we assign the variable(s) $\lambda$. These models are local in that $\lambda$ (at the very least) encodes all information in the intersection of the past light cones of both particles. If the position correlations can be explained by all local information $\lambda$, then the joint probability density $\rho(x_{1},x_{2})$ can be expressed in the following form:
\begin{equation}\label{LHVmodel}
\rho(x_{1},x_{2})=\int d\lambda\;\big(\rho(\lambda)\rho(x_{1}|\lambda)\rho(x_{2}|\lambda)\big).
\end{equation}
The CHSH and many other Bell inequalities, are mathematical consequences of LHV models of this form. By violating the CHSH inequality, we rule out the possibility that the joint probability density can be expressed in this form.
If we can show that the joint probability density is not expressible in this way, it then follows that knowing every piece of information in the past light cones of both particles cannot explain the correlations between them. This opens up new possibilities where there is either information outside both particles' light cones that will explain the correlations (implying a nonlocal universe), or that there is no information at all that will explain these correlations (implying a non-deterministic universe). As quantum metaphysics, these are interesting questions, but well beyond the scope of this work.

The derivation of the CHSH Bell inequality is based both on the essential LHV model, and on the measurements considered having a bounded (though possibly continuous) spectrum of possible results \footnote{Alternative derivations of Bell inequalities from the same LHV model include the CH74 inequality in which the measurement statistics are bounded differently than in the CHSH inequality.}. To keep things general, let $x$ be a real-valued random variable, and $f(x)$ be a function whose range is in the closed set $[-1,1]$. Let $\alpha$, and $\alpha'$ be two possible measurement settings for one party, and let $\beta$ and $\beta'$ be two possible measurement settings for a second party. If the position statistics can be described by an LHV model \eqref{LHVmodel}, then the correlation $\langle f(x_{1})f(x_{2})\rangle_{\alpha,\beta}$ must be expressible as the form:
\begin{equation}\label{LHVcorrModel}
\langle f(x_{1})f(x_{2})\rangle_{\alpha,\beta}=\int d\lambda\;\rho(\lambda)\langle f(x_{1})\rangle_{\alpha,\lambda}\langle f(x_{2})\rangle_{\beta,\lambda},
\end{equation}
for some $\lambda$. Next, since for all $x, f(x)\in[-1,1]$, all expectations $\langle f(x)\rangle$ must fall within the same range, independent of conditioning. With a little extra algebra (see \cite{CHSHbell1969}), one can show that the inequality:
\begin{align}\label{CVBellIneq}
|\langle &f(x_{1})f(x_{2})\rangle_{\alpha,\beta}-\langle f(x_{1})f(x_{2})\rangle_{\alpha,\beta'}|\nn\\
&\mp\big(\langle f(x_{1})f(x_{2})\rangle_{\alpha',\beta}+\langle f(x_{1})f(x_{2})\rangle_{\alpha',\beta'}\big)\leq 2,
\end{align}
must be valid if an LHV model of the form \eqref{LHVmodel} exists. Indeed, when $x$ has a binary spectrum (i.e., $\{-1,+1\}$), we see how the CHSH inequality as it applies to polarization measurements \cite{FreedmanCHSH1976, Aspect1981, AspectCHSH1982} is a special case of this more general formulation.

\subsection{Bell's Wavefunction and CHSH Violation}
What makes Bell nonlocality in continuous variables especially difficult is that many useful approximations to continuous-variable quantum states have positive-definite Wigner functions, and so explicitly admit an LHV model. To see how this works, we consider the Wigner function $W(x_{1},x_{2},k_{1},k_{2})$ for a pair of particles (in one spatial dimension) with position observables $\hat{x}_{1}$ and $\hat{x}_{2}$, and momentum observables $\hat{k}_{1}$ and $\hat{k}_{2}$. The joint probability density $\rho(x_{1},x_{2})$ can be expressed in terms of the Wigner function:
\begin{equation}
\rho(x_{1},x_{2})=\iint dk_{1}dk_{2} \;W(x_{1},x_{2},k_{1},k_{2}),
\end{equation}
where:
\begin{align}
W(&x_{1},x_{2},k_{1},k_{2})\equiv \frac{1}{(2\pi)^{2}}\iint dq_{1}dq_{2}e^{i(q_{1}x_{1}+q_{2}x_{2})}\nn\\
&\times\tilde{\psi}\Big(k_{1}+\frac{q_{1}}{2},k_{2}+\frac{q_{2}}{2}\Big)\tilde{\psi}^{\ast}\Big(k_{1}-\frac{q_{1}}{2},k_{2}-\frac{q_{2}}{2}\Big).
\end{align}
Note that if the correlations admit an LHV model, then $\rho(x_{1},x_{2})$ has the form seen in equation \eqref{LHVmodel}. Because of this, when the Wigner function happens to be positive-definite (obeying the form of a probability density), we can immediately describe $\rho(x_{1},x_{2})$ with an LHV model; the hidden variables $\lambda$ would be the arguments of the Wigner function, while the probability densities in parentheses would constitute the Wigner function itself.

\begin{figure*}[t]
 \centering
\includegraphics[width=\textwidth]{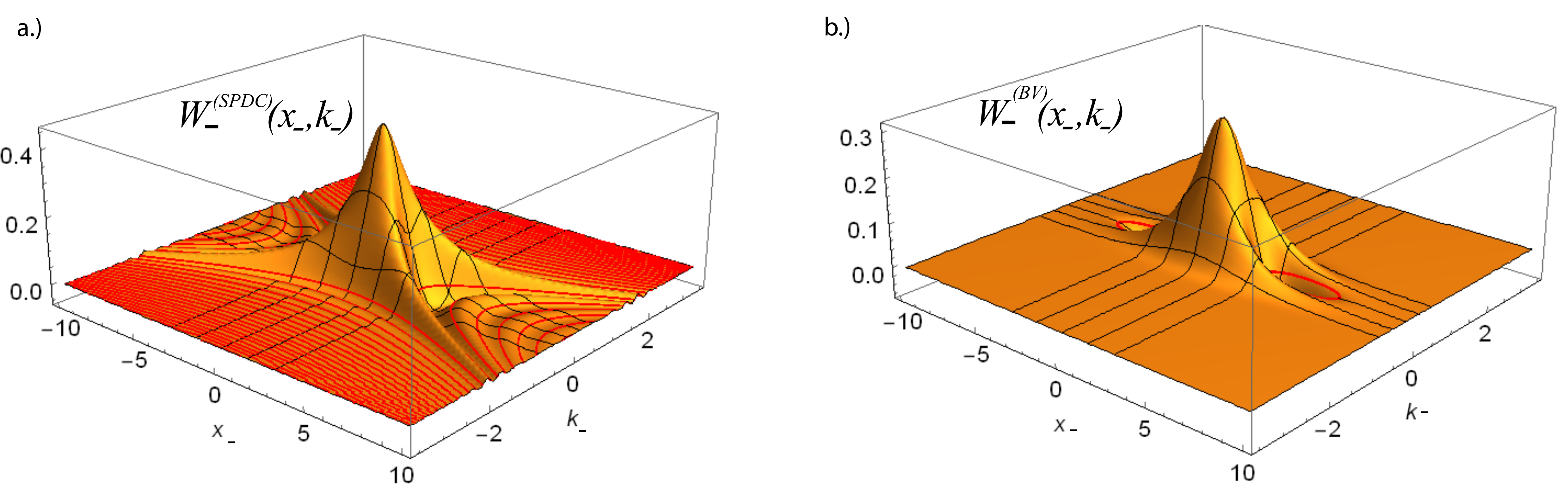}
\caption{(Color online) On the left (a) is a plot of the Wigner function $W_{-}^{(SPDC)}(x_{-},k_{-})$ obtained from direct calculations of the biphoton amplitude from Spontaneous Parametric Down-Conversion (SPDC), a popular source of entangled photon pairs (as shown in \cite{Schneeloch2015SPDC, LawEberly2004, MonkenSPDCPRA1998}). We note that although the biphoton Wigner function can be approximated as a Gaussian, there are significant regions of negativity. On the right (b) is a plot of Bell's Wigner function $W_{-}^{(BV)}(x_{-},k_{-})$ \eqref{BellWignerMinus} for $\sigma_{-}$ chosen to match position moments of the SPDC biphoton state. The values below the level (thick red) contours are negative.}
\end{figure*}

Since wavefunctions with non-negative Wigner functions admit an LHV model, we know that such states will never violate Bell inequalities based on \emph{those measurement statistics}. In spite of these difficulties, there are relatively simple wavefunctions that can violate a continuous-variable Bell inequality. Indeed, Bell provided a specific example in \cite{bell2004speakable}:
\begin{equation}\label{psibellCV}
\psi^{(BV)}(x_{1},x_{2})=N((x_{1}-x_{2})^{2} - 8\sigma_{-}^{2})e^{-\frac{(x_{1}+x_{2})^{2}}{8\sigma_{+}^{2}}}e^{-\frac{(x_{1}-x_{2})^{2}}{8\sigma_{-}^{2}}},
\end{equation}
where $N$ is a normalization constant. This ``Bell-Violating" wavefunction is especially convenient because it is separable in terms of the rotated (orthogonal) coordinates:
\begin{equation}
x_{+}\equiv \frac{x_{1}+x_{2}}{\sqrt{2}}\qquad,\qquad x_{-}\equiv \frac{x_{1}-x_{2}}{\sqrt{2}}.
\end{equation}
As a result, the Wigner function for $\psi^{(BV)}$ is separable this way as well, i.e.,
\begin{equation}
W^{(BV)}(x_{1},x_{2},k_{1},k_{2})=W^{(BV)}_{+}(x_{+},k_{+})W^{(BV)}_{-}(x_{-}k_{-}),
\end{equation} 
where
\begin{equation}
W(x_{+},k_{+})\equiv \frac{1}{2\pi}\iint dq_{+}e^{iq_{+}x_{+}}\tilde{\psi}\Big(k_{+}+\frac{q_{+}}{2}\Big)\tilde{\psi}^{\ast}\Big(k_{+}-\frac{q_{+}}{2}\Big),
\end{equation}
and $W(x_{-},k_{-})$ is similarly defined for $x_{-}$ and $k_{-}$.
Now, $W^{(BV)}_{+}(x_{+},k_{+})$ is a Gaussian:
\begin{equation}
W_{+}^{(BV)}(x_{+},k_{+})=\frac{1}{\pi}e^{-\frac{x_{+}^{2}}{4\sigma_{+}^{2}}}e^{-\frac{k_{+}^{2}}{4\big(\frac{1}{4\sigma_{+}}\big)^{2}}},
\end{equation}
but since $\psi^{(BV)}(x_{1},x_{2})$ has both quadratic and Gaussian factors depending on $x_{-}$, $W_{-}^{(BV)}(x_{-},k_{-})$ is a more elaborate function with negative values:
\begin{align}\label{BellWignerMinus}
&W_{-}^{(BV)}(x_{-},k_{-})=\frac{1}{11\pi \sigma_{-}^{4}}\Big(x_{-}^{4} + 2 x_{-}^{2}\sigma_{-}^{2}(-5+4 k_{-}^{2}\sigma_{-}^{2})\nn\\ 
&+ \sigma_{-}^{4}\big(11 + 8k_{-}^{2}\sigma_{-}^{2}+16k_{-}^{4}\sigma_{-}^{4}\big)\Big)e^{-\frac{x_{-}^{2}}{2\sigma_{-}^{2}}}e^{-\frac{k_{-}^{2}}{2(\frac{1}{4\sigma_{-}})^{2}}}.
\end{align}
In Fig.~1b, we plotted $W_{-}^{(BV)}(x_{-},k_{-})$ to show where the negative values occur.

To demonstrate Bell nonlocality for continuous variables (in this case, the positions $x_{1}$ and $x_{2}$ of a pair of particles), Bell examined (for different measurement settings) the signs of the position measurement outcomes ($+1$ for $x>0$, and $-1$ for $x\leq 0$). He then took for his correlation measure the mean of the product of these signs. Since these mean products are bounded between $-1$ and $1$, these statistics can be readily used in the CHSH inequality to test for nonlocality in continuous variables (e.g., $f(x)$ in \eqref{LHVcorrModel} could be Bell's piecewise sign function). To examine these correlations for different measurement settings, he treated the pair of systems described by the wavefunction \eqref{psibellCV} as though they were free non-interacting particles, and time evolved their joint state to different points in time $t_{1}$ and $t_{2}$ according to the free particle Schr\"{o}dinger equation. Calculating the sign correlation at different pairs of times allowed him to show by example how one might violate the CHSH Bell inequality with continuous-variable states.

\begin{figure*}[t]
 \centering
\includegraphics[width=\textwidth]{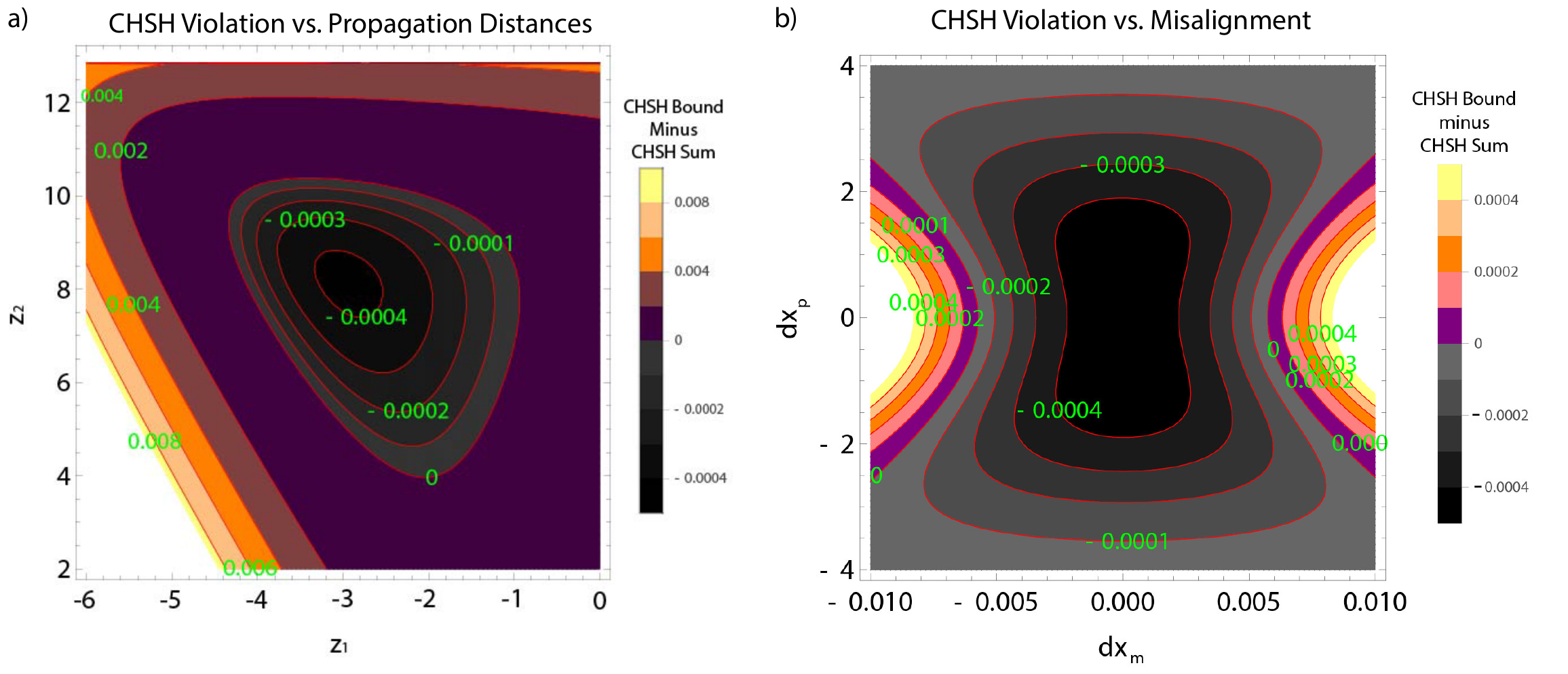}
\caption{(Color online) On the left (a) is a contour plot of the violation of the CHSH inequality for different propagation distances $z_{1}$ and $z_{2}$. These plots were calculated using Bell's wavefunction with $\sigma_{+}=1$mm, $\sigma_{-}=0.01$mm, and a wavelength of $650$nm for each photon. On the right (b) is a contour plot of the violation of the CHSH inequality at the optimal values of $z_{1}$ and $z_{2}$ ($-3$ and $+8$ mm respectively) for different misalignments of the detectors. $dx_{p}=dx_{1}+dx_{2}$ is a parallel position misalignment (both shifted the same direction), while $dx_{m}=dx_{1}-dx_{2}$ is an anti-parallel misalignment (both shifted in opposite directions). A negative value indicates a violation of the CHSH inequality.}
\end{figure*}

\section{Adapting Bell's Approach to Measurements of Photon Pairs}
Although Bell showed theoretical violation of the CHSH inequality by propagating his wavefunction with the free particle Hamiltonian to different times, one may violate the CHSH inequality experimentally by considering Bell's wavefunction as describing a biphoton transverse position amplitude (say, for signal and idler photons in degenerate spontaneous parametric down-conversion (SPDC)), and considering the sign correlation measurement performed at different propagation distances, as opposed to different times. The justification for this comes from the fact that the paraxial Helmholtz equation (governing the propagation of the individual photons) is mathematically identical (variable names aside) to the Schr{\"o}dinger equation for a free particle moving in two transverse dimensions, e.g.,
\begin{equation}\label{isomorphism}
-\frac{\partial^{2} A}{\partial x^{2}}-\frac{\partial^{2} A}{\partial y^{2}}=i k_{p}\frac{\partial A}{\partial z}\;\sim\; -\frac{\partial^{2} \Psi}{\partial x^{2}}-\frac{\partial^{2} \Psi}{\partial y^{2}}=i \frac{2m}{\hbar}\frac{\partial \Psi}{\partial t},
\end{equation}
where $A(x,y,z)$ gives the spatial dependence of the amplitude of either the signal or idler electric field; and $k_{p}$ is the wavenumber of the pump electric field (i.e, twice the signal or idler wavenumbers).

Using his Wigner function $W^{(BV)}(x_{1},k_{1};x_{2},k_{2})$, in the approximation that $\sigma_{+}\gg\sigma_{-}$ (so that $W^{(BV)}_{+}(x_{+},k_{+})$ may be neglected as a constant factor), Bell showed a qualitative violation of the CHSH inequality \eqref{CVBellIneq} for optimum measurement settings.  However, if we are to use this relationship \eqref{isomorphism} to get a \emph{quantitative} violation, we need to show that a violation is possible without such approximation, as the diffraction of light implies this approximation only works for a narrow range of distances. Indeed, in Bell's own treatment, the spreading of the entangled wavepacket over time implies his approximation (and subsequent result) is only valid for a narrow range of time settings \cite{Johansen1997}.

\subsection{Quantitative Violation of CHSH Inequality with Bell's Wavefunction as a Biphoton Amplitude}
In order to see whether it is empirically feasible to violate the CHSH inequality for position and momentum with sign binning, we need to be able to violate this inequality without taking such limits (e.g., $\sigma_{+}\gg\sigma_{-}$). By performing numerical calculations of the Fresnel-propagated biphoton field, we find that with a biphoton field resembling Bell's wavefunction, and using typical values of $\sigma_{+}=1$mm and $\sigma_{-}=0.01$mm, we were able to show that we could violate the CHSH inequality \eqref{CVBellIneq}, but only by at most a small amount (i.e., our maximum value was $2.00041 $, though with a numerical error bound less than $10^{-8}$) (see Fig.~2). In order to realize such a minute violation, we would need approximately $10^{8}$ coincidence counts to barely resolve each of the sixteen probabilities in the CHSH inequality to an uncertainty of $\pm 1.0\times 10^{-4}$, and that is assuming any systematic error due to misalignment (see Fig.~2b) or other factors is insignificant. We also performed similar calculations for the biphoton state created from type-1 SPDC, but any violation was inconclusive as the uncertainty in the results from these highly oscillatory probability densities was larger than the obtained violations. Although this underscores the difficulty in a successful demonstration of position-momentum Bell-nonlocality, it is still useful to consider what sort of experimental setup might be used in a successful demonstration once a more suitable state is discovered.

\subsection{Imagining an Experimental Demonstration with Photon Pairs}
The idealized setup we consider is as seen in Fig.~3. A pump laser and nonlinear crystal serve as a source of entangled photon pairs though other sources may be substituted. The pump light is filtered out, and the photon pairs are separated by a 50/50 beamsplitter into separate arms. Each arm contains a 4F imaging system, where spatially resolving pairs of photon detectors are placed in the image planes conjugate to the exit face of the photon pair source. In addition, these detectors are placed on translation stages, allowing us to measure the sign correlations for different propagation distances in the different arms. 

The reason we would need a 4F imaging system in each arm is that the biphoton field formed in those conjugate planes would be identical to the field just as it exits the source (though reflected and subject to the paraxial approximation). Without it, negative propagation distances would be impossible to realize, as there are no photon pairs before the source \footnote{That using a 4F imaging system resolves our issue with negative propagation distances came as a result of discussions with Prof. Miguel Alonso of the University of Rochester.}.

Though the 50/50 beamsplitter halves the collection efficiency of photon pairs, it does not alter the transverse spatial statistics of the photon pairs. Alternatively, if we use a source of entangled photon pairs of orthogonal polarizations (as with type-2 SPDC), a polarizing beamsplitter can better separate the pairs, allowing for a larger collection efficiency. Here, we are interested only in an idealized setup that may be improved upon in future developments.

\begin{figure}[t]
 \centering
\includegraphics[width=\columnwidth]{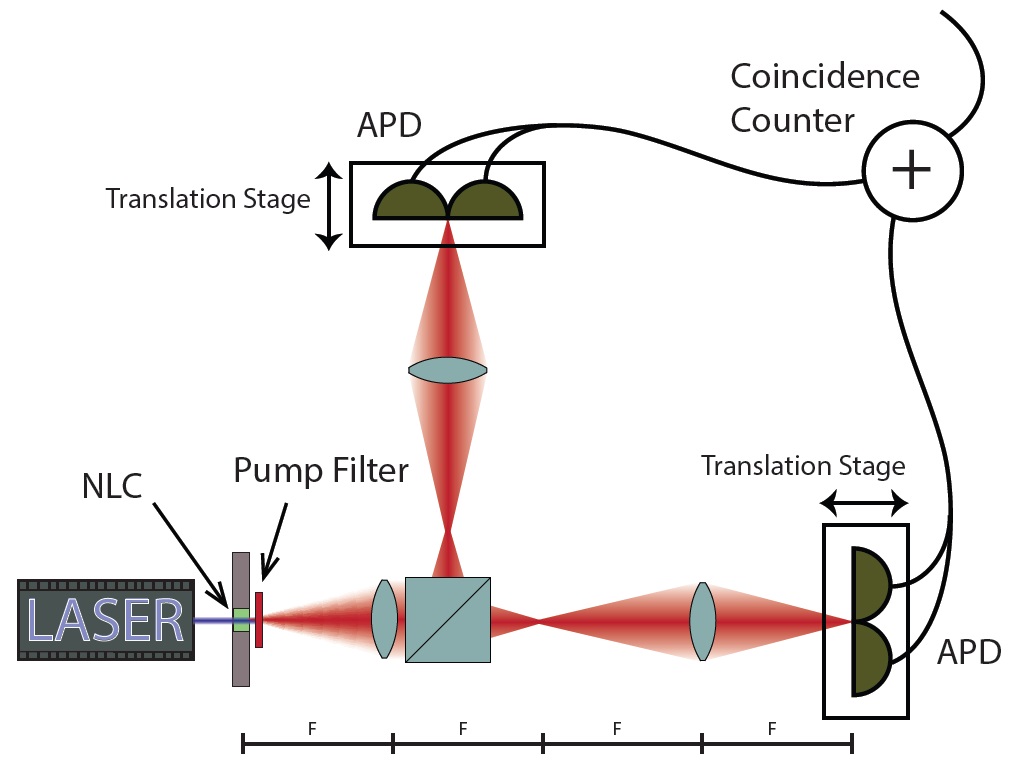}
\caption{(Color online) Here is a diagram of an idealized setup to violate the CHSH inequality for position and momentum with sign-binning. A laser and a nonlinear crystal (NLC) serve as a source of entangled photon pairs, whose state is determined by both the particulars of the laser and the geometry of the NLC. We would then image the exit face of this photon-pair source onto pairs of avalanche photo-diodes (APDs) with a 4F imaging system. By translating the APDs along the optic axis, we may use coincidence counts to measure the transverse sign correlations as a function of a variable propagation distance in each arm.}
\end{figure}

\subsection{Engineering a Suitable Biphoton Amplitude}
Once a more suitable biphoton amplitude is discovered, it is important to consider how such a state can be created experimentally. Here we consider engineering a biphoton state from a source of SPDC photon pairs. The biphoton amplitude in SPDC is determined by two major factors. The first factor is the pump spatial profile, which we may control with a spatial light modulator, and standard optical components. The second factor is how the second-order nonlinear coefficient varies over the length of the nonlinear crystal (what we call the longitudinal nonlinearity profile). For standard nonlinear crystals, this nonlinearity profile is a constant top-hat function over the propagation distance $z$ (being one value within the crystal, and zero outside the crystal). As discussed in \cite{DixonGaussPhaseMatchTailorOptExp2013}, the transverse biphoton amplitude (in momentum space) is related to the Fourier transform of the nonlinearity profile of the crystal. If we could continuously vary the nonlinearity within the nonlinear crystal, while controlling the pump spatial profile, we could exactly reproduce Bell's wavefunction among many others as a transverse spatial amplitude for photon pairs in SPDC.

Though we cannot \emph{continuously} vary the nonlinearity profile of the crystal, it is possible (as also shown in \cite{DixonGaussPhaseMatchTailorOptExp2013}) to use a periodically poled nonlinear crystal, and adjust the duty cycle (i.e., the fraction of positive to negative poling within each poling period) as a function of the crystal length to get a biphoton amplitude closely resembling the state we want. The high fidelity required for the approximating biphoton wavefunction in the periodically poled nonlinear crystal make this approach challenging, though not outside the realm of possibility.

\section{Conclusion}
Historically, the CHSH inequality and other Bell inequalities have largely been used for finite-dimensional discrete observables. However, Bell himself showed how one might use the CHSH inequality to witness the nonlocality of continuous observables. Since a local hidden variable model for a continuous-variable joint probability distribution implies a similar model for any statistics derived from that distribution, violating the CHSH inequality with sign binned statistics demonstrates the nonlocality of the underlying continuous observables. Although Bell showed a qualitative demonstration of position-momentum Bell-nonlocality, we showed that the state he considers gives a small quantitative violation, though other states may yet do much better. In addition, we showed how his approach can be re-envisioned as an experiment measuring similar correlations between entangled photon pairs, and discussed important challenges and issues, which when overcome, lead to a successful demonstration of position-momentum Bell-nonlocality.

\begin{acknowledgments}
We gratefully acknowledge useful discussions with Dr. Gregory Howland and Prof. Miguel Alonso as well as support from the National Research Council fellowship with the Air Force Research Laboratory In Rome, NY, and support from DARPA DSO InPho Grant No. W911NF-10-1-0404, DARPA DSO Grant No. W31P4Q-12-1-0015, and AFOSR Grant No. FA9550-13-1-0019.
\end{acknowledgments}

\bibliography{EPRbib12}

\end{document}